\documentclass[11pt]{article}
\usepackage{graphicx}
\usepackage{slashed}
\usepackage{amsmath,amssymb,amsfonts}
\usepackage{braket}
\usepackage[colorlinks=true, linkcolor=blue, citecolor=blue, linktoc=all]{hyperref}
\usepackage{comment,cite}

\usepackage[makeroom]{cancel}

\begin{document}
\newcommand{\uiuc}[1]{
	\centerline{
		\begin{minipage}[c]{0.7\linewidth}
			\begin{center}
			${}^{#1}$Illinois Center for Advanced Studies of the Universe \& Department of Physics,\\ 
			University of Illinois, 1110 West Green St., Urbana IL 61801, U.S.A.
			\end{center}
		\end{minipage}
		}
	}
\newcommand{\thistitle}{
	Complexity for link complement States in Chern Simons Theory 
	}
\newcommand{\emailrgl}{rgleigh@illinois.edu}

\title{\thistitle}
\author{Robert G. Leigh and Pin-Chun Pai\\
	\\
	{\small \emph{\uiuc{}}}\\ 
	\\
	}
\date{\today}
\maketitle
\begin{abstract}
We study notions of complexity for link complement states in Chern Simons theory with compact gauge group $G$. Such states are obtained by the Euclidean path integral on the complement of $n$-component links inside a 3-manifold $M_3$. For the Abelian theory at level $k$ we find that a natural set of fundamental gates exists and one can identify the complexity as differences of linking numbers modulo $k$. Such linking numbers can be viewed as coordinates which embeds all link complement states into $\mathbb{Z}_k ^{\otimes n(n-1)/2}$ and the complexity is identified as the distance with respect to a particular norm. For non-Abelian Chern Simons theories, the situation is much more complicated. We focus here on torus link states and show that the problem can be reduced to defining complexity for a single knot complement state. We suggest a systematic way to choose a set of minimal universal generators for single knot complement states and then evaluate the complexity  using such generators. A detailed illustration is shown for $SU(2)_k$ Chern Simons theory and the results can be extended to general compact gauge group.
\end{abstract}

%

\section{Introduction}
Quantum information concepts play an important role in high energy and gravitational research.
For example, in the context of holographic duality there are well-known calculational tools for entanglement entropy via the Ryu-Takayanagi formula and its extensions \cite{Ryu:2006bv,Engelhardt:2014gca,Rangamani:2016dms}. Other quantum information concepts such as complexity have also been conjectured to have dual gravitational interpretations\cite{Alishahiha:2015rta,Carmi:2016wjl,Stanford:2014jda,Susskind:2014moa,Brown:2015lvg,Brown:2015bva,Couch:2016exn}, although this is much less understood. Generally, the problem is that notions such as complexity are difficult to formulate in quantum systems with many degrees of freedom, such as quantum field theories or gravity. It is thus of interest to explore these concepts in simple field theories  (see for example \cite{Jefferson:2017sdb} for early attempts).

In the context of quantum circuits, circuit (or computational) complexity is defined as follows: given a reference state $\ket{\phi_R}$ and a set of fundamental gates, i.e., a set of unitary operators $\{\hat{\mathcal {U}}^I\}$, the complexity of a target state $\ket{\phi_T}$ is the minimum number $N$ of fundamental gates needed to map  $\ket{\phi_R}$ to $\ket{\phi_T}$,
\begin{eqnarray*}
\ket{\phi_T}= \hat{\mathcal {U}}^{i_1}\hat{\mathcal {U}}^{i_2}\cdots \hat{\mathcal {U}}^{i_N}\ket{\phi_R}.
\end{eqnarray*}
Clearly, there are many features of this description that imply that the notion of circuit complexity may be ambiguous. For example, the number and nature of the fundamental gates presumably matters. In simple qubit circuits, one often requires the gates to be "small",  involving only one or two qubits. As the quantum theory becomes more complicated (and generic quantum field theories are indeed complicated), these choices become more involved. 
 
In some situations, complexity may be addressed through
Nielsen's geometric approach \cite{Nielsen}, which identifies the complexity as the length of a geodesic on the space of operators spanned by such generators.
In this approach, for given reference state $\ket{\phi_R}$ and target state $\ket{\phi_T}$, one tries to find a unitary operator $\hat{U}$ such that
\begin{eqnarray}
\label{map}
\hat{U} \ket{\phi_R} = \ket{\phi_T} .
\end{eqnarray}
The operator $\hat{U}$ satisfying the above equation is not unique. One then expresses such operators in path integral form:
\begin{eqnarray}
\label{Upath}
\hat{U} = \mathcal{P} exp \left( i \int^1_0ds \sum_I Y_I(s)\hat{O}^I\right)
\end{eqnarray}
where $\{\hat{O}^I \}$ are a set of fundamental generators and $\{Y_I(s)\}$ describes a path from the identity operator to $\hat{U}$ in the space of operators. The next step is to define a cost $\mathcal{D}$ for each possible path 
\begin{eqnarray}
\label{cost}
\mathcal{D} = \int^1_0ds\ F\left(Y_I(s), \frac{d}{ds}Y_I(s)\right),
\end{eqnarray}
where $F$ is some local cost function. The cost of each path can be interpreted as  length in a Finsler geometry. The complexity of $\hat{U}$ is defined as the minimal cost of all possible paths satisfying \eqref{Upath}
\begin{eqnarray*}
\mathcal{C} _{\hat{U}} \equiv min \ \mathcal{D}.
\end{eqnarray*}
The geometric complexity of a target state is defined as the minimal complexity among all possible $\hat{U}$ satisfying \eqref{map}
\begin{eqnarray*}
\mathcal{C} _{\ket{\phi_T}} \equiv min \ \mathcal{C} _{\hat{U}}
\end{eqnarray*}
In \cite{Nielsen} it was shown that such a  definition can be related to circuit complexity 
up to a polynomial $poly(n)$ in $n$, where $n$ is the number of fundamental generators.\footnote{In Nielsen's original paper \cite{Nielsen}, $n$ was the number of qubits. }
 This approach gives a geometrical interpretation to computational complexity. The $poly(n)$ is some non-trivial polynomial factor depending on the set of generators, cost function and the target state. In most applications, such a factor is ignored and geometric complexity is taken as an estimation of circuit complexity. On the other hand, since the definition of circuit complexity in quantum field theory is unclear, geometric complexity may be viewed as a  definition of complexity. This approach associates complexity to a geometric object and can be potentially extended to continuous state space. However, the form of a cost function $F$ is still ambiguous, except in some models for which a natural candidate appears \cite{Caputa:2018kdj,Magan:2018nmu}. Some usual choices are the one-norm, i.e.,
\begin{eqnarray}
F_1 \equiv \sum_I  \left| \frac{d Y_I}{ds}  \right|
\end{eqnarray}
or two-norm
\begin{eqnarray}
F_2 \equiv  \sqrt{ \sum_I \left( \frac{d Y_I}{ds} \right)^2  }.
\end{eqnarray}
For example,  \cite{Jefferson:2017sdb} considered Gaussian states in free field theory, which can be generated by a finite set of generators. They test $\kappa $-norms for $\kappa \in \mathbb{R}^+$ as cost functions and conclude that complexity defined by $\kappa=1$ has the cutoff dependence most similar to wormhole volume \cite{Stanford:2014jda}.

There are on the other hand different proposals to define complexity in quantum field theories which do not rely on geometric methods. For example, the so-called "Path-Integral Complexity" is defined for CFT by minimizing certain functionals \cite{Caputa:2017urj,Caputa:2017yrh}. This approach also shows similar cutoff dependence structure to the volume or action \cite{Caputa:2017yrh},\footnote{However, the coefficients do not match in general.}
 and a recent study \cite{Erdmenger:2020sup} suggests that this path-integral complexity is related to geometric complexity in some models \cite{Caputa:2018kdj,Magan:2018nmu}. 
%
%

In this paper, we study the complexity in the context of Chern Simons theory. Explicitly we are interested in Chern Simons theory for compact gauge group $G$ with level $k$. We consider these theories on 3-manifolds $\mathcal{M}_{\mathcal{L}^n }$, which are link complements of $n$-component links in $S^3$. Such manifolds have disconnected boundaries, which we take to be $n$ linked tori. The Euclidean path integral  defines states in the tensor product of Hilbert space associated to each torus. These states provide a good stage to study complexity for two reasons. First, as a topological field theory, such Chern Simons theories have a Hilbert space of finite dimension. Furthermore, since the objects we are interested in are topological invariants, the complexity is reflected in topological properties, encoded in the coloured Jones polynomials.

As we will see, there is a natural way to define the fundamental gates for $U(1)_k$ Chern Simons theory so the complexity of link complement states is well defined. In this case the complexity is directly connected to the Gauss linking numbers between components of the link $\mathcal{L}^n$. This observation provides the first example that topological properties can manifest in complexity. Then we move on to non-Abelian Chern Simons theory, in which defining complexity is considerably more complicated. To attack the problem, we first focus on torus link states and use the fact that such states always have (at least in a certain framing) GHZ-like structure \cite{Balasubramanian:2018por}. This property allows us to reduce the problem to defining complexity for a single knot state. We  choose a minimal set of  fundamental generators by physical considerations, and then show that by using such generators a systematic algorithm to calculate the computational complexity can be constructed. Our work provides a different way to define complexity for link complement states from \cite{Camilo:2019bbl}, in which the authors defined ``topological complexity" as the minimal number of topological operations such as modular transformations and subsequently suggested that topological complexity provides an upper bound for circuit complexity.

The rest of this paper is organized as follows: In Sec. \ref{LCstates} we review the properties of link complement states in Chern Simons theory. We then discuss the framing ambiguity in Sec. \ref{Framing_factor}. In Sec. \ref{U1CS} we study the complexity for $U(1)_k$ Chern Simons theory and show that a natural definition of fundamental gates and complexity is available. In Sec. \ref{TLstates} we turn to discuss torus link states for non-Abelian Chern Simons. We review the result that such states have GHZ-like structure and show that by introducing $CNOT$ operators the problem is reduced to defining complexity of a single knot state. In Sec. \ref{Mingates} we investigate how to define a set of minimal, fundamental generators for a single knot by some physical considerations. In Sec. \ref{Minpath} and Sec. \ref{Extension} we compute the complexity as the minimal effort needed to prepare the target state using such fundamental generators. We then briefly discuss how these constructions can be extended to more general states by releasing the constraint of "small" generators in Sec. \ref{exten}. In Sec. \ref{Picture} detailed computations of complexity is illustrated for the simplest class of torus links. Finally summary and discussion are provided in Sec. \ref{Discussion}.
\section{Link complement states in Chern Simons theory}\label{LCstates}

\begin{figure} [h!]
\begin{center}
\includegraphics[width=200pt]{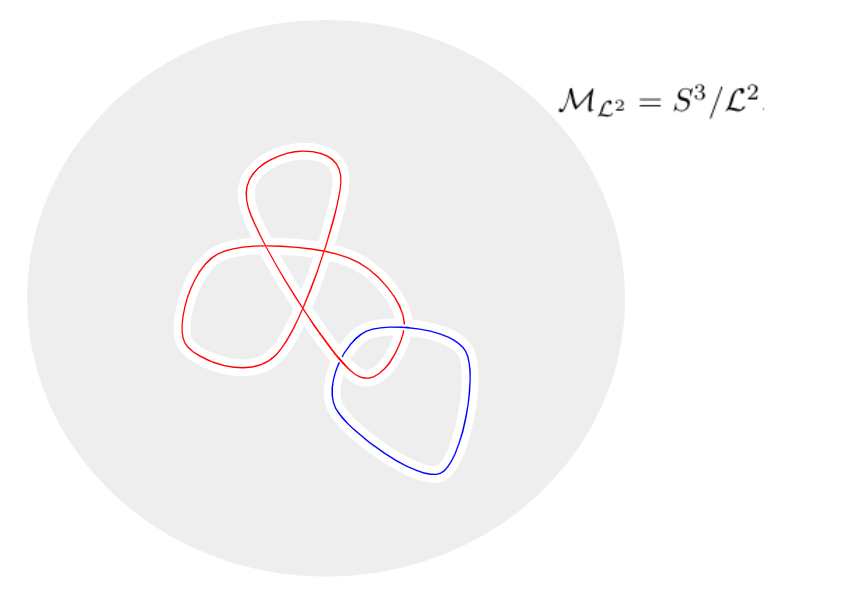} 
\caption{A case of $\mathcal{M}_{\mathcal{L}^2} = S^3/\mathcal{L}^2 $. To construct this manifold, one starts from $S^3$, represented by the total gray region, and then removes the tubular neighborhood of a link $\mathcal{L}^2 $. The resulting manifold has two disjoint torus boundaries, corresponding to the two components.}
\label{ML2}
\end{center}
\end{figure}

Consider Chern Simons theory on a closed 3-manifold $M$, which has the action
\begin{eqnarray*}
S_{CS}[A]=\frac{k}{4\pi} \int_M Tr(A\wedge dA+\frac{2}{3} A\wedge A\wedge A),
\end{eqnarray*}
where $A$ is a connection for a principal bundle on $M$ with structure group $G$  and $k$ is the (integer) level. The classical equation of motion is 
$F=dA + A \wedge A = 0,
$ which requires the connection to be flat. If $M$ has boundary $\Sigma$, the path integral on $M$ with boundary conditions $A|_{\Sigma} = A_0$,
\begin{eqnarray}
\label{path}
\Psi [ A_0] = \int_{A|_{\Sigma} = A_0} [DA] e^{iS_{CS}[A]}
\end{eqnarray}
is interpreted as the wavefunction of a state in the Hilbert space $\mathcal{H}(\Sigma;G,k)$ associated to $\Sigma$. As in \cite{Balasubramanian:2016sro,Balasubramanian:2018por}, we consider the case that $M$ is a link complement of the 3-sphere $S^3$, which is denoted by $\mathcal{M}_{\mathcal{L}^n }$. Such a manifold is constructed by first putting a non-self-intersecting $n$-component link, $\mathcal{L}^n = \sqcup_{i=1}^n L_i $ on $S^3$ and then removing the tubular neighborhood of the link from $S^3$ (see Fig. \ref{ML2} for an example). By this construction we have 
\begin{eqnarray*}
\Sigma_n = \partial\mathcal{M}_{\mathcal{L}^n } = \sqcup_{i=1}^n T^2.
\end{eqnarray*}
 \\
In other words, the boundary of $\mathcal{M}_{\mathcal{L}^n }$ is a disjoint union of $T^2$'s, which surround the $n$ components of $\mathcal{L}^n$. The Hilbert space is the $n$-fold tensor product $\mathcal{H}^{\otimes n}$, where $\mathcal{H} = \mathcal{H}(T^2;G,k)$ is the Hilbert space of Chern Simons theory for the group $G$ at level $k$ on each torus. Hence the path integral (\ref{path}) on $\mathcal{M}_{\mathcal{L}^n }$ associates the link $\mathcal{L}^n$ with a state in $\mathcal{H}^{\otimes n}$, which we will denote by $\ket{\mathcal{L}^n }$.  
We can expand such link complement states in the basis of $\mathcal{H}(T^2;G,k)$, $\{\ket{q}\}$. It is well known that for compact gauge group $G$, the Hilbert space on a torus $\mathcal{H}(T^2;G,k)$ has finite dimension, and the basis $\{\ket{q}\}$ is related one-to-one  to the irreducible representations $\mathcal{R}_q$ of $G$ with level $k$ \cite{Witten:1988hf}. Given a link $\mathcal{L}^n$ with $n$ components $\{L_1,L_2,\cdots ,L_n\}$, the associated link complement state $\ket{\mathcal{L}^n}$ can be expressed as
\begin{eqnarray}
\label{linkstate}
\ket{\mathcal{L}^n}&=& C_0 \sum_{q_1,q_2,...,q_n} \mathcal{C}_{\mathcal{L}^n}(q_1,q_2,...,q_n) \ket{q_1}_1 \otimes \ket{q_2}_2 \otimes\cdots \otimes\ket{q_n}_n \\  \nonumber
&\equiv& C_0 \sum_{q_1,q_2,...,q_n} \mathcal{C}_{\mathcal{L}^n}(q_1,q_2,...,q_n)  \ket{q_1,q_2 \cdots q_n} 
\end{eqnarray}
where $ \{\ket{q_i}_i \}$ is the basis of for the Hilbert space associated to the torus surrounding the $i^{th}$ component and $C_0$ is a normalization constant. The coefficients can be determined by surgery methods \cite{Witten:1988hf}: 
\begin{eqnarray*}
 \mathcal{C}_{\mathcal{L}^n}(q_1,q_2,...,q_n) = \braket{q_1,q_2 \cdots q_n|\mathcal{L}^n} = \braket{W_{R^\ast_{q_1}}(L_1) \cdots W_{R^\ast_{q_n}}(L_n)  }_{S^3}
\end{eqnarray*}
where $W_{\mathcal{R}_q}(L) = Tr_{\mathcal{R}_q} (e^{i \int_L A}) $ is the Wilson loop operator along knot $L$ in the representation $\mathcal{R}_q$. In other words, $ \mathcal{C}_{\mathcal{L}^n}(q_1,q_2,...,q_n)$ is the expectation value of Wilson loop operators along the link ${\mathcal{L}^n}$, also known as the  coloured link invariant. Therefore, the topological properties of the link ${\mathcal{L}^n}$ is encoded in its corresponding state $\ket{\mathcal{L}^n}$ through the coefficients, and we expect that complexity must also manifest these properties in some way.
 
In the remainder of this paper, we will investigate how to define complexity for such link complement states in various cases. To do this we need to choose a reference state first. It is natural to choose this to be  the simplest link, the $n$ component unknot, which we denote by $ \mathcal{L}_0^{\otimes n}$, i.e.
\begin{eqnarray*}
\ket{\phi_R} =  \ket{\mathcal{L}_0^{\otimes n}}.
\end{eqnarray*}

\subsection{Framing ambiguity }\label{Framing_factor}

Before studying complexity of link complement states, we should deal with the issue of framing \cite{Witten:1988hf} of each component comprising the link $\mathcal{L}^n$. The framing ambiguity can be thought of as self-linking, if we view each component as a ribbon instead of a circle. To fix the ambiguity, one must pick a framing for each component. If one chooses different framing such that the $i^{th}$ component varies by $t_i$ units, the link complement state will differ by $t_i$ Dehn twist on the corresponding basis, $\{\ket{q}_i\}$. Explicitly the states will transform as
\begin{eqnarray}
\ket{\mathcal{L}^n} \rightarrow \left( \mathcal{T}_1^{t_1} \otimes   \mathcal{T}_2^{t_2} \otimes \dots   \mathcal{T}_n^{t_n} \right) \ket{\mathcal{L}^n} \equiv \ket{\mathcal{L}^n}_D
\end{eqnarray} 
where $ \mathcal{T}_i$ is a Dehn-twist on the $i^{th}$ torus. However, since $\mathcal{T}_i$ are all unitary, such framing factors can be always absorbed by redefining gates as follows
\begin{eqnarray*}
 \hat{\mathcal {U}}^I  \rightarrow \left( \mathcal{T}_1^{t_1} \otimes   \mathcal{T}_2^{t_2} \otimes \dots   \mathcal{T}_n^{t_n} \right)  \hat{\mathcal {U}}^I 
\left( \mathcal{T}_1^{t_1} \otimes   \mathcal{T}_2^{t_2} \otimes \dots   \mathcal{T}_n^{t_n} \right) ^{-1} \equiv  \hat{\mathcal {U}}^I_D .
\end{eqnarray*} 
Given a combination of gates which can map the reference state to the target state
\begin{eqnarray*}
\ket{\mathcal{L}^n}  = \hat{\mathcal {U}}^{i_1}\hat{\mathcal {U}}^{i_2}\cdots \hat{\mathcal {U}}^{i_N} \ket{\mathcal{L}_0^{\otimes n}},
\end{eqnarray*}
one can immediately find the corresponding circuit which has the same feature under the different choice of framing
\begin{eqnarray*}
\ket{\mathcal{L}^n}_D  = \hat{\mathcal {U}}^{i_1}_D\hat{\mathcal {U}}^{i_2}_D\cdots \hat{\mathcal {U}}^{i_N}_D \ket{\mathcal{L}_0^{\otimes n}}_D,
\end{eqnarray*}
so any reasonable definition of complexity remains unchanged. In the case of the geometric approach the situation is similar. One can redefine generators $\{\hat{O}^I\}$ in the same way to create the corresponding circuit under different framing, without changing the path in operator space, $Y_I(s)$. It again implies that the complexity is invariant. Therefore the framing factor can  always be absorbed into the choice of gates or generators and complexity can be defined without ambiguity.
\section{Complexity for link complement states}
\subsection{$U(1)_k$ case}\label{U1CS} \label{U1CS} 

We start with the simplest Abelian case, in which the gauge group is $U(1)$ with level $k$. It is well-known that $\mathcal{H}(T^2;U(1),k)$ is $k$-dimensional. Furthermore, the coloured Jones polynomials for $U(1)_k$ Chern Simons only depend on the linking numbers between components of ${\mathcal{L}^n}$. Denoting the linking number between $L_a$ and $L_b$ by $l_{ab}$, the normalized link complement state is \cite{Witten:1988hf}
\begin{eqnarray*}
\ket{\mathcal{L}^n}&=& \frac{1}{k^{n/2}} \sum_{q_1,q_2,...,q_n} exp \left(  \frac{2 \pi i}{k}\sum_{a < b} l_{ab} q_a q_b  \right) \ket{q_1,q_2 \cdots q_n}. 
\end{eqnarray*}
where $q_i =0, 1, \cdots,k-1$. In this section we will denote the link complement state as $\ket{\mathcal{L}^n(l_{ab})}$ since it only depends on the linking numbers. We can view the $\frac{n(n-1)}{2}$ linking numbers $\{l_{ab}\}$ as ``coordinates" on the space of $n$-link complement states $\mathcal{H}_{\mathcal{L}^n}$. It is clear that these coordinates have the periodicity $l_{ab} \sim l_{ab}+k $. Therefore $\mathcal{H}_{\mathcal{L}^n}$ is a discrete, compact space, with the topology 
\begin{eqnarray*}
\mathcal{H}_{\mathcal{L}^n}  = \mathbb{Z}_k ^{\otimes n(n-1)/2}.
\end{eqnarray*}
Since the components of $ \mathcal{L}_0^{\otimes n}$ do not wind around each other, the reference state is simply 
\begin{eqnarray*}
\ket{\phi_R} =  \ket{\mathcal{L}_0^{\otimes n}}= \frac{1}{k^{n/2}} \sum_{q_1,q_2,...,q_n}  \ket{q_1,q_2 \cdots q_n} 
\end{eqnarray*}
which we take as the `origin' of $\mathcal{H}_{\mathcal{L}^n}$. The evolution from the reference state $ \ket{\mathcal{L}_0^{\otimes n}}$ to the target state $\ket{\mathcal{L}^n(l_{ab})}$ can be done by the following unitary operation:
\begin{eqnarray}
\label{Uab}
\ket{\mathcal{L}^n(l_{ab})}=    exp \left(  \frac{2 \pi i}{k}\sum_{a < b} l_{ab} \hat{q}_a \otimes \hat{q}_b  \right) \ket{\mathcal{L}_0^{\otimes n}} 
\end{eqnarray}
where $\hat{q}_a$ is the local operator acting on $a^{th}$ site as 
\begin{eqnarray*}
\hat{q}_a \ket{q_1,q_2 \cdots q_n}  = q_a \ket{q_1,q_2 \cdots q_n}.
\end{eqnarray*}
The mapping (\ref{Uab}) can be viewed as a quantum circuit and it can be prepared by a finite number of fundamental gates, which we define as 
\begin{eqnarray}
 \hat{\mathcal{U}}_{ab} \equiv exp \left( \frac{2 \pi i}{k} \hat{q}_a \otimes \hat{q}_b \right )
\end{eqnarray}
and their inverses. These are the analogue of 2 qubit gates. In terms of these gates we can write the operator in (\ref{Uab}) as
\begin{eqnarray}
 exp \left(  \frac{2 \pi i}{k}\sum_{a < b} l_{ab} \hat{q}_a \otimes \hat{q}_b  \right) = \prod_{a<b} (\hat{\mathcal{U}}_{ab})^{l_{ab}}
\end{eqnarray}
Therefore we  act with each $\hat{\mathcal{U}}_{ab}$  $(l_{ab} \ mod \ k)$ times on the reference state to obtain the target state. If $(l_{ab} \ mod \ k) > \frac{k}{2}$ then we should use $(\hat{\mathcal{U}}_{ab}^{-1})^{k -(l_{ab} \ mod \ k) }$ instead of $(\hat{\mathcal{U}}_{ab})^{l_{ab}}$ to get a shorter circuit. The complexity of a link complement state is then defined as
\begin{eqnarray}
\label{complex}
\mathcal{C}(\ket{\mathcal{L}^n(l_{ab})})  = \sum_{a<b}  min [ (l_{ab} \ mod \ k),   k -(l_{ab} \ mod \ k)].
\end{eqnarray} 
The meaning of the fundamental gates is clear: acting with $\hat{\mathcal{U}}_{ab}$ once is equivalent to adding one linking number between $L_a$ and $L_b$, i.e. $l_{ab} \rightarrow l_{ab} +1$, while $\hat{\mathcal{U}}_{ab}^{-1}$ corresponds to decreasing the linking number by one. In other words, the gates are unit translation operators in $\mathcal{H}_{\mathcal{L}^n}$. The complexity (\ref{complex}) turns out to be the minimal number of steps needed to translate from $\ket{ \mathcal{L}_0^{\otimes n}}$ to $\ket{\mathcal{L}^n(l_{ab})}$.

So far we have considered the case that the reference state is $\ket{ \mathcal{L}_0^{\otimes n}}$, which is the origin of $\mathcal{H}_{\mathcal{L}^n}$. However, $\ket{ \mathcal{L}_0^{\otimes n}}$ plays no special role in the above discussion. We are free to set any link complement state $\ket{\mathcal{L}^n(l_{ab}^\prime)}$ as the reference state. Since
\begin{eqnarray*}
 exp \left(  \frac{2 \pi i}{k}\sum_{a < b} l_{ab} \hat{q}_a \otimes \hat{q}_b  \right) \ket{\mathcal{L}^{n }(l^\prime_{ab})}= \ket{\mathcal{L}^{n}(l^\prime_{ab}+l_{ab})},
\end{eqnarray*}
$\ket{ \mathcal{L}_0^{\otimes n}}$ is just the special case that $l^\prime_{ab} = 0$. The complexity with the general reference state is now defined as 
\begin{eqnarray*}
\mathcal{C}(\ket{\mathcal{L}^n(l_{ab})})_{\ket{\phi_R} = \ket{\mathcal{L}^n(l_{ab}^\prime)}}  &=& \sum_{a<b}  min [ (\Delta l_{ab} \ mod \ k),   k -(\Delta l_{ab} \ mod \ k)]  \\
\Delta l_{ab} &\equiv&| l_{ab} - l_{ab}^\prime |.
\end{eqnarray*} 
We see that the complexity of link complement states in $U(1)_k$ Chern Simons is naturally defined. In this case all link complement states form a discrete, compact space $\mathbb{Z}_k ^{\otimes n(n-1)/2}$. The number of fundamental gates is the same as the dimension of $\mathbb{Z}_k ^{\otimes n(n-1)/2}$, which is sensible because we can span the state space by such gates. The complexity is identified as the distance with respect to the $l_1$-norm on this space.

%
As expected, the complexity of a link complement state is related to the link's topological properties. In this simplest Abelian case we see that the fundamental gates just correspond to increasing or decreasing Gauss linking numbers between components of a link, which are sufficient to determine the corresponding state in $U(1)_k$ Chern Simons. An important property of such gates is that they are all commuting so the structure of complexity is so simple that one can separate the contribution from each gate. In the next section we will find that for non-Abelian gauge group the situation is much more complicated so we need to use a different strategy to define the complexity.

\subsection{$SU(2)_k$ link complement states}

For non-Abelian Chern Simons it is harder to define complexity because the link complement states are more involved. Recall that a link complement state 
\begin{eqnarray}
\ket{\mathcal{L}^n}= C_0 \sum_{q_1,q_2,...,q_n} \mathcal{C}_{\mathcal{L}^n}(q_1,q_2,...,q_n)  \ket{q_1,q_2 \cdots q_n} 
\end{eqnarray}
is determined by the coloured link invariants. Unlike the Abelian case, coloured link invariants for non-Abelian Chern Simons depend on more than the Gauss linking numbers. There is no simple analytic form for arbitrary link complement states.\footnote{In principle one can calculate the coloured Jones polynomials by braiding operations \cite{Kaul:1991np,Kaul:1993hb,Kaul:1998ye}, but the expression apparently cannot be written as a quantum circuit acting on some reference state.}
To simplify the question, we will restrict the target states considered to be a subset of all link complement states. If the subset has some symmetry property, then one can use fewer parameters to describe it so defining complexity could be simpler. We will focus on torus link complement states in the remainder of this paper, and leave further generalizations to future work. The topological complexity of torus knots was studied in \cite{Camilo:2019bbl}, which was defined as the minimal number of modular transformations required to map an unknot to the target torus knot. It was then argued that topological complexity could be related to the circuit complexity of the corresponding state. However, given a modular transformation which maps a torus knot to another, in general the corresponding modular matrix is not a map between the corresponding knot complement states. Therefore, we do not expect torus knot complement states to form a representation of the modular group, $SL(2,\mathbb{Z})$. This fact prevents the identification of topological complexity of knots with the circuit complexity of their complement states. Here we provide a different approach to define the circuit complexity by directly investigating torus link complement states. Such links are highly symmetric and as we will see, the fusion rule and Verlinde formula allow us to reduce the problem to defining the complexity of states corresponding to single knots. 

\subsubsection{GHZ-like structure of torus link complement states} \label{TLstates}
 
A torus link with $n$ components can be labeled by $(nP,nQ)$, where $P$, $Q$ are two co-prime integers. We first recall that all torus link states have GHZ-like structure \cite{Balasubramanian:2018por}. Such states can be expressed as \cite{Isidro:1992fz,Labastida:2000yw,Brini:2011wi}
\begin{eqnarray}
\label{toruslink}
\ket{\mathcal{L}^n (nP,nQ)} = C_0 \sum_{q_1, \cdots, q_n}  J_{q_1, \cdots, q_n} (\mathcal{L}^n)\ket{q_1, \cdots, q_n}
\end{eqnarray} 
where $J_{q_1, \cdots, q_n} (\mathcal{L}^n)$ is the coloured Jones polynomial for the torus link and $C_0$ is a normalization constant. The coefficients can be represented in the following form:
\begin{eqnarray*}
 J_{q_1, \cdots, q_n} (\mathcal{L}^n) = \sum_{j_1, \cdots, j_{n-1}} N_{q_1 q_2 j_1} N_{j_1 q_3 j_2} \cdots N_{j_{n-2} q_n j_{n-1}} J_{j_{n-1}}(P,Q)
\end{eqnarray*} 
where $N_{ijk}$ are the fusion coefficients and $J_{j}(P,Q)$ are the coloured Jones polynomials of the $(P,Q)$ torus knot. Although $J_{j}(P,Q)$ can be computed in terms of modular matrices \cite{Labastida:1990bt}, we will not need it here. Using the Verlinde formula \cite{Verlinde:1988sn}
\begin{eqnarray*}
N_{ijk} = \sum_l  \frac{S_{il}S_{jl}S_{kl}}{S_{0l}}
\end{eqnarray*} 
where $S$ is the modular transformation matrix implementing $\tau \rightarrow  -1/\tau$ along with the fact that $S=S^T$ and $S^2 =1$, we arrive at 
\begin{eqnarray*}
 J_{q_1, \cdots, q_n} (\mathcal{L}^n) &=& \sum_{l_1,\cdots l_{n-1}}   \sum_{j_1,\cdots j_{n-1}}  
 \frac{S_{q_1l_1}S_{q_2l_1}S_{j_1l_1}}{S_{0l_1}}  \frac{S_{j_1l_2}S_{q_3l_2}S_{j_2l_2}}{S_{0l_2}} \cdots  \frac{S_{j_{n-2}l_{n-1}}S_{q_nl_{n-1}}S_{j_{n-1}l_{n-1}}}{S_{0l_{n-1}}} J_{j_{n-1}}(P,Q)  \\ &=&\sum_{l} \sum_{j_s}
\frac{S_{lq_1}S_{lq_2}\cdots S_{lq_n}  }{(S_{0l})^{n-1}}  S_{lj_s} J_{j_s}(P,Q)
\end{eqnarray*} 
Therefore (\ref{toruslink}) becomes
\begin{eqnarray}
\ket{\mathcal{L}^n (nP,nQ)} = C_0 \sum_{q_1, \cdots, q_n}   \sum_{l} \sum_{j_s} \frac{S_{lq_1}S_{lq_2}\cdots S_{lq_n} }{(S_{0l})^{n-1}} S_{lj_s} J_{j_s}(P,Q)\ket{q_1, \cdots, q_n}
\end{eqnarray}
The above form can be further simplified if we (unitarily) change the basis to $\ket{\tilde{l}} \equiv S_{lj} \ket{j}$: 
\begin{eqnarray}
\label{GHZ}
\ket{\mathcal{L}^n (nP,nQ)} =  \sum_{l} f_{l}\ket{\tilde{l}, \cdots, \tilde{l}}
\end{eqnarray}
where
\begin{eqnarray}
f_{l} \equiv C_0 \sum_{j_s} \frac{1}{(S_{0l})^{n-1}} S_{lj_s} J_{j_s}(P,Q).
\end{eqnarray}
Obviously (\ref{GHZ}) has the GHZ-like structure.\footnote{This GHZ-like structure manifests for a specific choice of framing. We admit such framing as a natural one since this structure reflects the symmetry of torus links clearly. 
}
Let us also write the reference state $\ket{ \mathcal{L}_0^{\otimes n}}$ in the new basis:
\begin{eqnarray*}
\ket{ \mathcal{L}_0^{\otimes n}} &=&  \sum_{q_1, \cdots, q_n}  S_{0q_1}S_{0q_2}\cdots S_{0q_n}\ket{q_1, \cdots, q_n}  \\
&=& \ket{\tilde{0}, \tilde{0}, \cdots, \tilde{0}}.
\end{eqnarray*}
One can see that the reference state takes a simple form in the $\ket{\tilde{q}}$ basis. It is convenient to introduce the so-called Controlled-NOT gates (CNOT). In the context of quantum computation, CNOT is a quantum logic gate involving two qubits, which operates as
\begin{eqnarray*}
\hat{C}_{NOT} \ket{a} \otimes \ket{b} = \ket{a} \otimes \ket{(a+b)\  mod \ 2}
\end{eqnarray*}
where $a,b$ = $0,1$. Here we define similar gates, denoted as $\hat{C}_{NOT}^{ij}$, which act on the $i^{th}$ and $j^{th}$ sites as (assuming $i<j$):
\begin{eqnarray*}
\hat{C}_{NOT}^{ij} \ket{\tilde{q_1}} \otimes\cdots\ket{\tilde{q_i}}\cdots\ket{\tilde{q_j}} \cdots \otimes\ket{\tilde{q_n}} = \ket{\tilde{q_1}} \otimes\cdots\ket{\tilde{q_i}}\cdots\ket{(\tilde{q}_i+\tilde{q}_j) \ mod \ D} \cdots \otimes\ket{\tilde{q_n}} 
\end{eqnarray*}
where $D$ is the dimension of the Hilbert space for a single site, i.e. $q_i=0,1,\cdots,D-1$. Therefore if we start from a state $\ket{\phi^n}$ which has the following form 
\begin{eqnarray*}
\ket{\phi^n} = \sum_q f_q \ket{ \tilde{q},\tilde{0}, \tilde{0}, \cdots, \tilde{0}},
\end{eqnarray*}
the CNOT gates act on it as
\begin{eqnarray*}
\hat{C}_{NOT}^{12} \ket{\phi^n} &=& \sum_q f_q \ket{ \tilde{q},\tilde{q}, \tilde{0}, \tilde{0}, \tilde{0}, \cdots, \tilde{0}}  \\
\hat{C}_{NOT}^{13} \hat{C}_{NOT}^{12} \ket{\phi^n} &=& \sum_q f_q \ket{ \tilde{q},\tilde{q}, \tilde{q}, \tilde{0}, \tilde{0}, \cdots, \tilde{0}} \\
&\cdot& \\ &\cdot& \\ &\cdot& \\
\hat{C}_{NOT}^{1n} \cdots \hat{C}_{NOT}^{13} \hat{C}_{NOT}^{12} \ket{\phi^n} &=& \sum_q f_q \ket{ \tilde{q},\tilde{q}, \tilde{q}, \tilde{q}, \tilde{q}, \cdots, \tilde{q}} \\
\end{eqnarray*}
In other words, once we know how to prepare
\begin{eqnarray}
\sum_{q} f_{q}\ket{\tilde{q},\tilde{0},\tilde{0}, \cdots, \tilde{0}}
\end{eqnarray}
from the reference state $\ket{ \tilde{0},\tilde{0}, \tilde{0}, \cdots, \tilde{0}}$, then the torus link state can be obtained by further acting with  $n-1$ CNOT gates. The problem reduces to defining the complexity of the state on a single site, i.e. the state corresponding to a single knot. The total complexity is then the sum $\mathcal{C}_{tot} = \mathcal{C}_{single\  knot} +  \mathcal{C}_{CNOT}$. 

\subsubsection{Universal, minimal generators for single knot states} \label{Mingates}

In this section, we discuss how to define the complexity for a single knot state, $\mathcal{C}_{knot}$. This is equivalent to defining the complexity for a target state $\ket{\phi_T} = \sum_q f_{q} \ket {\tilde{q}}$, given the reference state $\ket{\phi_R} = \ket {\tilde{0}}$. For simplification we will focus on $SU(2)_k$ Chern Simons theory in the rest of this paper but the result can be extended to general compact gauge groups. It is known that $\mathcal{H}(T^2;SU(2),k)$ has dimension $k+1$, so $q = 0,1,2,\cdots k$. Let us begin by considering the geometric approach \cite{Nielsen}. In this approach for given reference state $\ket{\phi_R}$ and target state $\ket{\phi_T}$ in $\mathbb{CP}^{k+1}$ one looks for an operator $\hat{U}\in SU(k+1)$ such that
\begin{eqnarray}
\ket{\phi_T} = \hat{U} \ket{\phi_R}.
\end{eqnarray}
One then writes such operators in integral form:
\begin{eqnarray}
\hat{U} = \mathcal{P} exp \left( i \int^1_0 \sum_I Y_I(s)  \hat{O}^I  \ ds \right)
\end{eqnarray}
In the geometric approach, $\{\hat{O}^I \}$ are supposed to be all the `small' generators, which here means all generators of $SU(k+1)$ since we are considering a single site. A cost $\mathcal{D}$ is defined for each possible path 
\begin{eqnarray}
\mathcal{D} = \int^1_0 F\left(  Y_I(s) , \frac{d}{ds}  Y_I(s)   \right) \  ds ,
\end{eqnarray}
where $F$ is some local cost function. Taking  inspiration from quantum circuits, it is natural to consider the $l_1$-norm,
\begin{eqnarray}
 F_1 \equiv \sum_I \left| \frac{d}{ds}  Y_I(s) \right|.
\end{eqnarray}
However, this choice breaks the homogeneity of $SU(k+1)$ so using different bases of generators will give different answers. On the other hand, if one uses the $l_2$-norm as the cost function so that the homogeneity is preserved then the resulting complexity is simply the `angle' separating the reference and target states. This is unsatisfying since we expect complexity to involve more features of the target state. It is still possible to consider a more complicated local function which also respects the homogeneity, but the relation to computational complexity becomes ambiguous. 

A possible solution is to break the homogeneity by some physical considerations so that one can obtain a natural, non-trivial definition of complexity. For example, in \cite{Brown:2019whu} the authors study the complexity of a single qubit, and they set the cost of $\sigma_z$ to be smaller than $\sigma_x$ and $\sigma_y$ by assuming that some laboratory design, such as magnetic field $B_z$, makes rotation in the $z$-direction easier. More precisely, the cost function used in \cite{Brown:2019whu} is
\begin{eqnarray*}
 F = \sqrt{ I_{xx} \left( \frac{d}{ds}  Y_x(s) \right)^2+I_{yy} \left( \frac{d}{ds}  Y_y(s) \right)^2 +I_{zz} \left( \frac{d}{ds}  Y_z(s) \right)^2  }.
\end{eqnarray*}
where $I_{xx} = I_{yy} \ll I_{zz}$.
 
In this paper we will use a different strategy to break the homogeneity. We take inspiration from the previous analysis of the Abelian case. Recall that for $U(1)_k$ Chern Simons we find the number of fundamental gates is $n(n-1)/2$, which equals  the dimension of the state space $\mathbb{Z}_k ^{\otimes n(n-1)/2}$. This makes sense because the minimal number of generators to span a space is equal to its dimension. Similarly, since we are considering states in $\mathbb{CP}^{k+1}$, which has $2k$ degrees of freedom, we may expect the minimal number of generators needed to prepare arbitrary target states is also $2k$. If we construct the quantum circuit by only using $2k$ generators instead of all generators of $SU(k+1)$, the homogeneity will be broken and one can define a non-trivial complexity.

Breaking homogeneity means, as we mentioned above, that some basis of Hilbert space plays a special role. From the above discussion we see that $\{\ket {\tilde{q}}\}$ is a natural candidate for three reasons. First of all, one of the basis elements, $\ket{\tilde{0}}$, corresponds to the reference state directly. Secondly, the GHZ-like structure of torus link complement states manifests in this basis. Finally, $\{ \ket{\tilde{q}}\}$ is a  natural choice in the context of Chern Simons theory since they are defined by Wilson loop operators and modular transformation, while arbitrary linear combinations do not have clear physical meaning in general.
 
Once we understand that $\{ \ket{\tilde{q}}\}$ is a special basis in the analysis, we can introduce $2k$ universal generators based on it to define the complexity. By universal we mean that such generators are sufficient to generate all states in $\mathbb{CP}^{k+1}$. Since we start from the reference state, the $2k$ generators can be chosen as rotation generators from $\ket{\tilde{0}}$ to $\ket{\tilde{i}}$, where $i = 1,2,\cdots, k$. To illustrate how to choose such generators, let us consider the simplest case $k=1$, in which the map from reference state to target state can be written as
\begin{eqnarray*}
 \left (
\begin{array}{ccc}
1\\
0
\end{array} 
\right)  \overset{\hat{\mathcal{O}}_1}{\rightarrow}
 \left (
\begin{array}{ccc}
f_0\\
f_1
\end{array} 
\right)
\end{eqnarray*}
where $\hat{\mathcal{O}}_1 \in SU(2)$. If we choose the phase such that $f_0$ is real then the general form of $\hat{\mathcal{O}}_1$ is 
\begin{eqnarray*}
\hat{\mathcal{O}}_1 = 
 \left (
\begin{array}{ccc}
f_0 & f_1^\ast \\
f_1 & f_0^\ast
\end{array} 
\right)
= \left (
\begin{array}{ccc}
cos \theta & sin \theta\ e^{-i \phi} \\
sin \theta\ e^{i \phi} & cos \theta
\end{array} 
\right)
\end{eqnarray*}
where we express $(f_0, f_1)$ in terms of two new variables $(\theta,\phi)$. 
We would like to express the above form using two universal generators. A naive choice may be $(\sigma_z,\sigma_y)$ so that
\begin{eqnarray*}
\hat{\mathcal{O}}_1 = e^{ \frac{i\phi}{2} \sigma_z } e^{ i \theta \sigma_y } e^{ -\frac{i\phi}{2} \sigma_z }.
\end{eqnarray*}
According to the above form, we might suggest that the complexity of $\hat{\mathcal{O}}_1$ could be defined as
\begin{eqnarray*}
\mathcal{C} (\hat{\mathcal{O}}_1 ) \propto |\theta| + |\phi|
\end{eqnarray*}
as the amount needed for each generator. However, the contribution of the phase term, $|\phi|$, is ambiguous because it depends on the choice of generators. For example, if one had decided to use $(\sigma_z,\sigma_x)$ as fundamental generators instead, the net result would be to replace the phase term by $|\phi - \frac{\pi}{2}|$,  but we do not have a good reason to distinguish the two choices. Actually, since $ \sigma_z $ does not change the reference state, the more natural choice of generators seems to be $(\sigma_x,\sigma_y)$. The operator can then be written as
\begin{eqnarray*}
\hat{\mathcal{O}}_1 = exp \left[ i \theta ( -sin \phi \ \sigma_x  + cos \phi \  \sigma_y ) \right] \equiv  exp \left[ i ( n_x \sigma_x  + n_y  \sigma_y ) \right]
\end{eqnarray*}
The above is not of the conventional form for a quantum circuit. We expect that the contributions from $\sigma_x$ and $\sigma_y$ should be symmetric since they are not distinguishable by the reference state. Therefore we are led to define the complexity as
\begin{eqnarray*}
\mathcal{C} (\hat{\mathcal{O}}_1 ) \propto \sqrt{n_x^2+n_y^2} = |\theta|.
\end{eqnarray*}
Such a definition is invariant if we choose another basis $(\sigma_x^\prime,\sigma_y^\prime)$ on the $x-y$ plane, and the phase term $\phi$ is absorbed into the rotation symmetry. One can interpret this to mean that the minimal path simply goes in the right direction towards the target state, traversing the straight path from.

So far we have just reproduced the usual angle between two rays in a Hilbert space:
\begin{eqnarray*}
|\theta| =cos^{-1} f_0=cos^{-1} \left( \sqrt{|\braket{\phi_R|\phi_T}|^2} \right) 
\end{eqnarray*}
but this analysis will be helpful in finding a non-trivial definition of complexity for larger $k$. 

The next case we will consider is $k=2$, in which the reference state and target state have the following form:
\begin{eqnarray}\label{k=2case}
\ket{\phi_R} =
 \left (
\begin{array}{ccc}
1\\
0 \\
0
\end{array} 
\right)   \quad , \quad  \ket{\phi_T} =
 \left (
\begin{array}{ccc}
f_0\\
f_1 \\
f_2
\end{array} 
\right).
\end{eqnarray}
Similar to the previous case, we seek four universal generators and a systematic process to map $\ket{\phi_R}$ to $\ket{\phi_T}$. One such process is composed of  two steps:
\begin{eqnarray}
\label{proce}
 \left (
\begin{array}{ccc}
1\\
0 \\
0
\end{array} 
\right)     \overset{\hat{\mathcal{U}}^{01}}{\rightarrow}
 \left (
\begin{array}{ccc}
\sqrt{1-|f_1|^2}\\
f_1 \\
0
\end{array} 
\right)
\overset{\hat{\mathcal{U}}^{02}}{\rightarrow}
 \left (
\begin{array}{ccc}
f_0\\
f_1 \\
f_2
\end{array} 
\right)
\end{eqnarray}
where we again set $f_0$  real. The above process corresponds to two $SU(2)$ rotations $\hat{\mathcal{U}}^{01}$ and $\hat{\mathcal{U}}^{02}$, analogous to $\hat{\mathcal{O}}_1 $ in the $k=1$ case. The minimal number of generators of them is four, as expected. The explicit forms of the two rotations are determined by the coefficients of the target state:
\begin{eqnarray}
\label{parameter}
\hat{\mathcal{U}}^{01} &=& 
 \left (
\begin{array}{ccc}
\sqrt{1-|f_1|^2} & f_1^\ast &0\\
f_1 & \sqrt{1-|f_1|^2} &0 \\
0&0&1
\end{array} 
\right)
\equiv \left (
\begin{array}{ccc}
cos \theta_1 & sin \theta_1 e^{-i \phi_1}  &0\\
sin \theta_1 e^{i \phi_1} & cos \theta_1 &0 \\
0 & 0 & 1
\end{array} 
\right) \\ \nonumber 
\hat{\mathcal{U}}^{02} &=& 
 \left (
\begin{array}{ccc}
f_0/ \sqrt{1-|f_1|^2} & 0 &  f_2^\ast / \sqrt{1-|f_1|^2}\\
0&1&0 \\
 f_2 / \sqrt{1-|f_1|^2} &0  & f_0 / \sqrt{1-|f_1|^2} 
\end{array} 
\right)
\equiv \left (
\begin{array}{ccc}
cos \theta_2 &0 & sin \theta_2 e^{-i \phi_2}  \\
0 & 1 & 0 \\
sin \theta_2 e^{i \phi_2} & 0& cos \theta_2  
\end{array} 
\right) 
\end{eqnarray}
where we reparametrize the coefficients by $(\theta_1,\phi_1,\theta_2,\phi_2)$. Now it is natural to define the complexity of this process as 
\begin{eqnarray*}
\mathcal{C} (\hat{\mathcal{U}}^{01}  ) + \mathcal{C} (\hat{\mathcal{U}}^{02}  )  \propto  |\theta_1| + |\theta_2|.
\end{eqnarray*}
Notice that the above form is not the usual inner-product distance in the Hilbert space. The reason is that in our construction the two rotations $\hat{\mathcal{U}}^{01}$ and $\hat{\mathcal{U}}^{02}$ are distinguishable: mixing of their generators are not allowed. This distinction makes sense because we are using a physically meaningful basis.  

In the above discussion we see that the generators of $\hat{\mathcal{U}}^{01}$ and $\hat{\mathcal{U}}^{02}$ are universal because we can always find at least one circuit of the form (\ref{proce}) mapping the reference state to an arbitrary target state. However, given these generators the process (\ref{proce}) may not be the process of minimal complexity. In the next subsection we will discuss how to find the minimal process.

\subsubsection{The path of minimal complexity}\label{Minpath}

We have chosen the minimal universal $2k$ generators corresponding to the rotations between the reference state and basis $\{ \ket{\tilde{q}}\}$. Here we  discuss how to construct a circuit of minimal complexity using  these generators. For simplicity we start with the $k=2$ case and then extend the result to higher $k$. 

\begin{figure} [h!]
\begin{center}
\includegraphics[width=160pt]{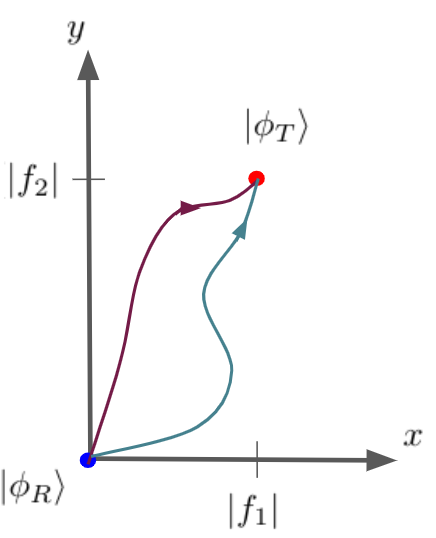} 
\caption{The reference state and target state on the $xy$ plane, and two possible paths connecting them.}
\label{process}
\end{center}
\end{figure}

For $k=2$ the reference state and the target state take the  form \eqref{k=2case} in the basis $\{\ket{\tilde{q}}\}$. The four fundamental generators are
\begin{eqnarray}
\label{generators}
\hat{\sigma}_x^{01}  \equiv  
 \left (
\begin{array}{ccc}
0 & 1 &0\\
1 & 0 &0 \\
0&0& 0
\end{array} 
\right)
\quad &\,& \quad 
\hat{\sigma}_y^{01}  \equiv  
 \left (
\begin{array}{ccc}
0 & i &0\\
-i & 0 &0 \\
0&0&  0
\end{array} 
\right)  \\  \nonumber
\hat{\sigma}_x^{02}  \equiv  
 \left (
\begin{array}{ccc}
0 & 0 &1\\
0 & 0&0 \\
1&0&0
\end{array} 
\right)
\quad &\,& \quad 
\hat{\sigma}_y^{02}  \equiv  
 \left (
\begin{array}{ccc}
0 & 0 &i\\
0 & 0 &0 \\
-i&0&0
\end{array} 
\right)
\end{eqnarray}
$\{ \hat{\sigma}^{0j}_{x,y}\}$ generate $SU(2)$ rotations between $\ket{\tilde{0}}$ and $\ket{\tilde{j}}$. As discussed in the last subsection, one expects that the phase of the $j^{th}$ coefficient, $arg(f_j)$, can be absorbed into the rotation symmetry of $\{ \hat{\sigma}^{0j}_{x,y}\}$. Therefore the complexity only depends on the magnitudes of coefficients, $\{|f_i|\}$, and the problem reduces to finding the following process of  minimal complexity:
\begin{eqnarray}
 \left (
\begin{array}{ccc}
1\\
0 \\
0
\end{array} 
\right)     \overset{P(s)}{\rightarrow}
\left (
\begin{array}{ccc}
\sqrt{1-|f_1|^2-|f_2|^2}\\
|f_1| \\
|f_2|
\end{array} 
\right)
\end{eqnarray}
by using only $\hat{\sigma}_y^{01}$ and $\hat{\sigma}_y^{02}$. A possible process $P(s)$ is described by a path on the first quadrant of the unit sphere parametrized by $0 \le s \le 1$, on which states are denoted by 
\begin{eqnarray}
\ket{\phi(s)} &=& \left (
\begin{array}{ccc}
\sqrt{1-x(s)^2-y(s)^2}\\
x(s) \\
y(s)
\end{array} 
\right)
\end{eqnarray}
with the boundary conditions (see Fig. \ref{process})
\begin{eqnarray}
\label{bc}
x(0) &= 0 \quad \,  \quad  & y(0) = 0  \\  \nonumber
x(1) &= |f_1| \quad \, \quad &  y(1)= |f_2|  
\end{eqnarray}
Now we want to compute the complexity of a given path. Consider the infinitesimal segment from $(x(s),y(s))$ to $(x(s+ds),y(s+ds))$; the starting point and endpoint are 
\begin{eqnarray}
\label{com1} \nonumber
\left (
\begin{array}{ccc}
\sqrt{1-x(s)^2-y(s)^2}\\
x(s) \\
y(s)
\end{array} 
\right)   &\rightarrow&
\left (
\begin{array}{ccc}
\sqrt{1-x(s+ds)^2-y(s+ds)^2}\\
x(s+ds) \\
y(s+ds)
\end{array} 
\right)
\\&&=
\left (
\begin{array}{ccc}
\sqrt{1-x(s)^2-y(s)^2}-\frac{ x \dot{x} ds  + y\dot{y} ds }{\sqrt{1-x(s)^2-y(s)^2}}\\
x(s) + \dot{x} ds \\
y(s) + \dot{y} ds
\end{array} 
\right) + O(ds^2)
\end{eqnarray}
This segment can be implemented by fundamental generators (\ref{generators}) via two infinitesimal angles $d \theta_1$ and $d \theta_2$: 
\begin{eqnarray*}
exp (i \sigma^{01}_y  d\theta_1)   &=&  \left (
\begin{array}{ccc}
1 & - d\theta_1 & 0 \\
d\theta_1 & 1 &0 \\
0&0&1
\end{array} 
\right)   + O(d\theta_1^2)  \\
exp (i \sigma^{02}_y  d\theta_2)   &=&  \left (
\begin{array}{ccc}
1 & 0 & -d\theta_2 \\
0 & 1 &0 \\
d\theta_2&0&1
\end{array} 
\right)   + O(d\theta_2^2)
\end{eqnarray*}
They map the starting point to 
\begin{eqnarray}
\label{com2}
exp (i \sigma^{01}_y  d\theta_1) \  exp (i \sigma^{02}_y  d\theta_2)  \left (
\begin{array}{ccc}
\sqrt{1-x(s)^2-y(s)^2}\\
x(s) \\
y(s)
\end{array} 
\right) =\left (
\begin{array}{ccc}
\sqrt{1-x(s)^2-y(s)^2}- (x d\theta_1 +y d \theta_2)\\
x(s) +\sqrt{1-x(s)^2-y(s)^2} d\theta_1 \\
y(s) + \sqrt{1-x(s)^2-y(s)^2} d\theta_2
\end{array} 
\right) + O(d\theta^2)
\end{eqnarray}
The order in (\ref{com2}) does not matter since their commutator is higher order. Comparing (\ref{com1}) and (\ref{com2}) we get
\begin{eqnarray*}
d\theta_1 &=&  \frac{ \dot{x} ds}{\sqrt{1-x(s)^2-y(s)^2}}  \\
d\theta_2 &=&   \frac{ \dot{y} ds}{\sqrt{1-x(s)^2-y(s)^2}} 
\end{eqnarray*}
Hence the complexity of this infinitesimal segment is naturally defined as
\begin{eqnarray}
d \mathcal{C}  \equiv |d\theta_1| +|d\theta_2| =  \frac{ (|\dot{x}| + |\dot{y}| ) ds}{\sqrt{1-x(s)^2-y(s)^2}}.
\end{eqnarray}
Thus we can compute the complexity of any process $P(s)$ through the functional 
\begin{eqnarray}
\mathcal{C} [P(s)] =\int d \mathcal{C} = \int_{P(s)}ds  \frac{ |\dot{x}| + |\dot{y}|  }{\sqrt{1-x(s)^2-y(s)^2}} .
\end{eqnarray}
By the variational method, one can solve for the path of minimal complexity for the given boundary conditions (\ref{bc}). Without loss of generality let us assume that $|f_1| \le |f_2|$. To simplify the problem, first consider a subset of all possible paths, $\mathcal{F}_+$, defined as  
 
$\bold{Definition\ 1: \mathcal{F}_+  =  \{ P(s) |\  \forall  s \in [0,1], \ \dot{x}(s) \ge 0, \ \dot{y}(s) \ge 0;\  satisfying\  boundary\  conditions\  (\ref{bc})\}}$  

It is easy to compute the Euler-Lagrange equations in $\mathcal{F}_+$, and later on we will see that the path of minimal complexity is contained in this subset. One can immediately recognize the following property:
   
$\bold{Property\ 1: \ if\  P(s) \in \mathcal{F}_+, \ then \ \forall (x,y) \ on \ P(s), \ x\le |f_1| \ and \ y \le |f_2|}$. 
 
\begin{figure} [h!]
\begin{center}
\includegraphics[width=450pt]{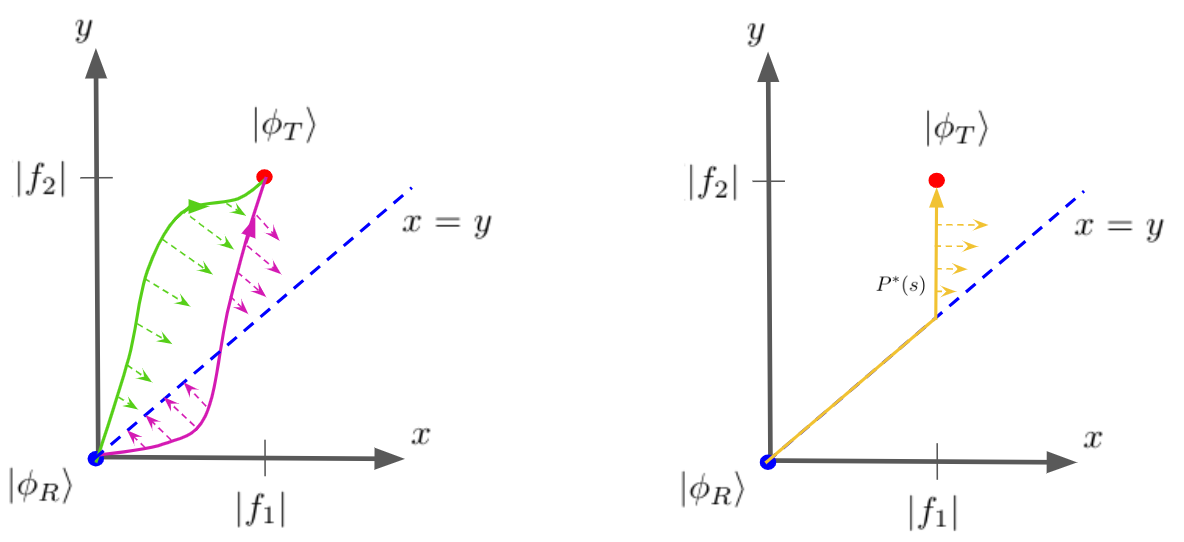} 
\caption{Left: two paths in $\mathcal{F}_+$. The dash arrows indicates the deviation flow to smaller complexity, which is given by (\ref{diff}). Notice that the flow always point towrd to $x=y$. Right: The path $P^\ast(s)$. Any deviation through the flow will give a new path no longer in $\mathcal{F}_+$ because it violates \textbf{Property1}, so $P^\ast(s)$ is the path closest to $x=y$ in $\mathcal{F}_+$.}
\label{compare}
\end{center}
\end{figure}

The complexity of paths in this subset is given by
\begin{eqnarray}
\mathcal{C} [P(s) \in  \mathcal{F}_+ ] = \int_0^1  \frac{ \dot{x} + \dot{y}  }{\sqrt{1-x^2-y^2}} ds   \equiv \int_0^1 L(x,y,\dot{x},\dot{y}) ds
\end{eqnarray}
and the Euler-Lagrange equations are
\begin{eqnarray*}
\frac{d}{ds}\frac{\partial L}{\partial \dot{x}} - \frac{\partial L}{\partial {x}} = 0 &\Rightarrow & \frac{(y-x)\dot{y}}{(1-x^2-y^2)^{3/2}} = 0  \\
\frac{d}{ds}\frac{\partial L}{\partial \dot{y}} - \frac{\partial L}{\partial {y}} = 0 &\Rightarrow & \frac{(x-y)\dot{x}}{(1-x^2-y^2)^{3/2}} = 0   
\end{eqnarray*}
These equations suggest that the minimal path must be $x=y$. However, this does not in general satisfy the boundary conditions (\ref{bc}), so we have to consider the path in $\mathcal{F}_+$ which is closest to $x=y$. Such a path is constructed piece-wise as (see Fig. \ref{compare})
\begin{eqnarray}
P^\ast(s) := \left\{\begin{array}{ll}
x(s) = y(s) = 2s| f_1|   &, \  \mbox{ $0 \le s \le \frac{1}{2}$}   \\  
x = |f_1|, \    y(s) = | f_1| + 2(|f_2|-|f_1|)(s-\frac{1}{2})   & , \ \mbox{$\frac{1}{2} \le s \le 1$}
                \end{array} \right.\label{Pstar}
\end{eqnarray}
The above is  composed of two straight lines on the $x$-$y$ plane: from $(0,0)$ to $(|f_1|,|f_1|)$, and then to $(|f_1|,|f_2|)$. Note that the segment with $x=const.$ is a solution of the reduced one-dimensional problem for $y(s)$, but we will elaborate on this in what follows.
Now we would like to prove the following:
  
$\bold{Claim\ 1:  P^\ast (s) \  is \ the \ path \ of\  minimal\  complexity\  in\  \mathcal{F}_+.}$ 
 
\begin{figure} [h!]
\begin{center}
\includegraphics[width=150pt]{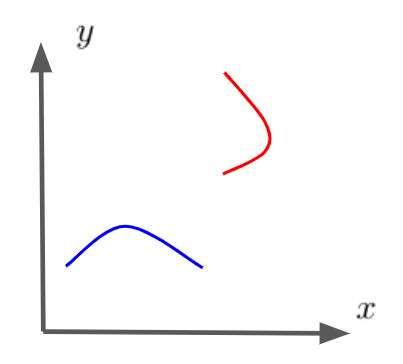} 
\caption{Examples of turnback segments of \textbf{(A)} (left curve) and \textbf{(B)} (right curve). }
\label{TB}
\end{center}
\end{figure}

We can use the variational method to prove this claim. Consider two nearby paths in $\mathcal{F}_+$, $P(s)$ and $P(s)+ \delta P(s)$. The difference in complexity is 
\begin{eqnarray}
\label{diff}
\delta \mathcal{C} = \int_0^1  \left[ \frac{(x-y)\dot{y}}{(1-x^2-y^2)^{3/2}} \delta x(s) + \frac{(y-x)\dot{x}}{(1-x^2-y^2)^{3/2}}\delta y(s)  \right] ds + O(\delta x, \delta y)^2
\end{eqnarray}
where we have used integration by parts and the fact that $(\delta x, \delta y)$ vanish at endpoints. Since we are considering the case $\dot{x}, \dot{y} \ge 0$, (\ref{diff}) suggests that we deform the path toward the following direction:
\begin{eqnarray}
sign(\delta x) = -sign(\delta y) = sign(y-x)
\end{eqnarray}
to get smaller complexity. Such direction always points toward $x=y$ (See Fig. \ref{compare}). 
Hence, for paths in $\mathcal{F}_+$, the closer to $x=y$, the smaller complexity we get. Since $P^\ast (s)$ is the path closest to $x=y$ in  $\mathcal{F}_+$, we have established $\bold{Claim\ 1}$.   
The next step is to prove that $P^\ast (s)$ is actually the one of minimal complexity among all possible paths. To do this, first define the concept of a ``turnback segment":
  
\textbf{Definition\ 2: A turnback segment is a path parametrized by $s \in [s_i ,s_f]  \subset [0,1]$, satisfying one of the following conditions (see Fig. \ref{TB}):\\(A) $y(s_i) = y(s_f)$ and $\forall s  \in (s_i ,s_f), y(s)>y(s_i).$   
\\(B) $x(s_i) = x(s_f)$ and $\forall s  \in (s_i ,s_f), x(s)>x(s_i).$ }   
\\

We will also use the following property:\\
\textbf{Property\ 2: If $P^\prime(s)$ is a path satisfying the boundary conditions (\ref{bc}) but is not in $\mathcal{F}_+$, then $P^\prime(s)$ contains at least one turnback segment.} 
 
\begin{figure} [h!]
\begin{center}
\includegraphics[width=400pt]{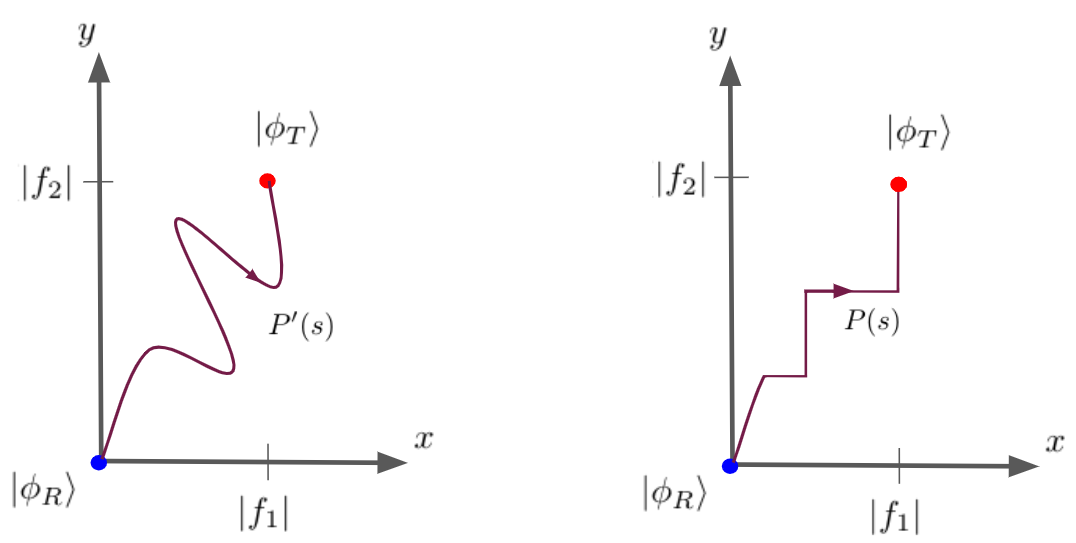} 
\caption{Left: A path $P^\prime(s)$ satisfying the boundary conditions (\ref{bc}) but not in $\mathcal{F}_+$. Such a path contains at least one turnback segment. Right: By replacing all turnback segments by vertical or horizontal lines, one can find a new path $P(s) \in\mathcal{F}_+$ having smaller complexity than $P^\prime(s)$.}
\label{C2}
\end{center}
\end{figure}

Now one can prove that for a path containing a turnback segment of type $\bold{(A)}$, we can always find another path with smaller complexity via replacing this  segment by a horizontal one, $ y=y(s_i)$. For such a turnback segment, we have
\begin{eqnarray}
\label{replace}
\int_{s_i}^{s_f}  \frac{ |\dot{x}| + |\dot{y}|  }{\sqrt{1-x(s)^2-y(s)^2}} ds \  >  \int_{s_i}^{s_f}   \frac{  |\dot{x}|  }{\sqrt{1-x(s)^2-y(s_i)^2}} ds
\end{eqnarray}
by using the definition of the type $\bold{(A)}$ turnback segment. The l.h.s of (\ref{replace}) is the complexity of the original segment and the r.h.s. is for the horizontal segment, so we prove the above statement. Similarly, if a path contains a turnback segment of type $\bold{(B)}$ then one can replace it by a vertical segment to get a new path with smaller complexity. Therefore, say we start from a path $P^\prime(s)$ satisfying the boundary conditions (\ref{bc}) but which is not in $\mathcal{F}_+$; by replacing all of its turnback segments by horizontal or vertical segments we get a new path $P(s) \in \mathcal{F}_+$ with smaller complexity. Thus we prove the following:
  
\textbf{Claim\ 2: For any $P^\prime(s)$ satisfying the boundary conditions (\ref{bc}) but not in $\mathcal{F}_+$, one can find a path $P(s) \in\mathcal{F}_+$ which has smaller complexity (See Fig. \ref{C2}).} \\
  \\ 
Combining \textbf{Claim\ 1} and \textbf{Claim\ 2} we conclude that $P^\ast(s)$ is the path of smallest complexity. Therefore the complexity of the target state $\ket{\phi_T}$ with respect to the reference state is, given \eqref{Pstar}, 
\begin{eqnarray*}
\mathcal{C}(\ket{\phi_T}) &=& \int_{P^\ast(s)}   \frac{ \dot{x} + \dot{y}  }{\sqrt{1-x(s)^2-y(s)^2}} ds  \\
&=&  \int^{1/2}_0    \frac{ 4|f_1 | }{\sqrt{1-8s^2|f_1|^2}} \  ds +   \int^{1/2}_0    \frac{ 2|f_2|-2|f_1 | }{\sqrt{1-|f_1|^2- (|f_1|-2s^\prime|f_1|+2s^\prime|f_2|)^2}} \  ds^\prime \\
&=& \sqrt{2}\  tan^{-1} \left( \frac{\sqrt{2} \ |f_1|}{\sqrt{1-2 |f_1|^2}} \right) + 
 tan^{-1} \left( \frac{ |f_2|}{\sqrt{1- |f_1|^2-|f_2|^2}} \right)
-  tan^{-1} \left( \frac{ |f_1|}{\sqrt{1-2 |f_1|^2}} \right)
\end{eqnarray*}
The first term arises from the first integral and is $\sqrt{2}$ times the angle between $\ket{\phi_R}$ and $(\sqrt{1-2|f_1|^2},|f_1|,|f_1|)$. The second integral gives the angle between $(\sqrt{1-2|f_1|^2},|f_1|,|f_1|)$ and $(|f_0|,|f_1|,|f_2|)$.
 For general target states, we can define $m \equiv min(|f_1|,|f_2|)$ and $M \equiv max(|f_1|,|f_2|)$. Then the complexity is defined as
\begin{eqnarray*}
\mathcal{C}(\ket{\phi_T}) 
= \sqrt{2}\  tan^{-1} \left( \frac{\sqrt{2} \ m}{\sqrt{1-2 m^2}} \right) + 
 tan^{-1} \left( \frac{ M}{\sqrt{1- m^2-M^2}} \right)
-  tan^{-1} \left( \frac{ m}{\sqrt{1-2 m^2}} \right).
\end{eqnarray*}
 
We pause here to make a comparison between our definition of complexity and Nielsen's geometric approach. One sees a similarity since in both cases, complexity is identified as some extremal functional. So, our definition can be viewed as a notion of geometric complexity. The difference is that we do not compute geodesic length in the whole space of unitary operators, but in a subspace. For example, in the $k=2$ case, we consider unitary operators which can be expressed as 
\begin{eqnarray*}
\hat{U} = P exp \left[ i \int^1_0 \left( Y_{1x}(s) \hat{\sigma}_x^{01} + Y_{1y}(s) \hat{\sigma}_y^{01} + Y_{2x}(s) \hat{\sigma}_x^{02}+Y_{2y}(s) \hat{\sigma}_y^{02} \right) ds \right],
\end{eqnarray*}
i.e., they can be generated by $4$ fundamental generators. Such operators only form a subspace of $SU(3)$ but they are sufficient to map the reference state to an arbitrary target state in $\mathbb{CP}^{3}$. The effective cost function in our approach is 
\begin{eqnarray*}
F = \sqrt{\dot{Y}_{1x}^2 + \dot{Y}_{1y}^2}+\sqrt{\dot{Y}_{2x}^2 + \dot{Y}_{2y}^2}.
\end{eqnarray*} 
The square root form ensures that the phases $arg(f_j)$, can be absorbed into the rotation symmetry of $\{ \hat{\sigma}^{0j}_{x,y}\}$.

\subsubsection{Extension to higher $k$} \label{Extension} 

For $k=3$ we have 
\begin{eqnarray*}
\ket{\phi_R} =
 \left (
\begin{array}{ccc}
1\\
0 \\
0 \\
0
\end{array} 
\right)   \quad , \quad  \ket{\phi_T} =
 \left (
\begin{array}{ccc}
\sqrt{1-|f_1|^2-|f_2|^2-|f_3|^2}\\
f_1 \\
f_2 \\
f_3
\end{array} 
\right).
\end{eqnarray*}
Similar to the $k=2$ case, we seek a path $P^\ast(s)$ on the first quadrant of the unit $3$-sphere which satisfies boundary conditions
\begin{eqnarray}
\label{bc2}
x(0) = 0 \quad \,  \quad  & y(0) = 0  \quad \,  \quad  & z(0) = 0  \\  \nonumber
x(1) = |f_1| \quad \, \quad &  y(1)= |f_2|   \quad \,  \quad  & z(1) = |f_3| 
\end{eqnarray}
and minimizes the complexity functional
\begin{eqnarray}
\mathcal{C} [P(s)] = \int_{P(s)}  \frac{ |\dot{x}| + |\dot{y}|+ |\dot{z}|  }{\sqrt{1-x(s)^2-y(s)^2 - z(s)^2}} ds .
\end{eqnarray}
We can define a set of paths, $\mathcal{F}_+$, analogous to \textbf{Definition\ 1} and show that $P^\ast(s) \in \mathcal{F}_+$. The Euler-Lagrangian equations for $\mathcal{F}_+$ are 
\begin{eqnarray*}
\frac{(y-x)\dot{y}+(z-x)\dot{z}}{(1-x^2-y^2-z^2)^{3/2}} &=& 0  \\
\frac{(x-y)\dot{x}+(z-y)\dot{z}}{(1-x^2-y^2-z^2)^{3/2}} &=& 0   \\
\frac{(x-z)\dot{x}+(y-z)\dot{y}}{(1-x^2-y^2-z^2)^{3/2}} &=& 0
\end{eqnarray*}
These equations can be satisfied if 
\begin{eqnarray}
\label{eom2}
x=y=z   .
\end{eqnarray}
Again, in general, (\ref{eom2}) does not satisfy the boundary conditions (\ref{bc2}). The minimal path $P^\ast(s)$ is the one in $\mathcal{F}_+$ which is closest to $x=y=z$. If $|f_1|\le|f_2|\le|f_3|$ then it takes the following form 
\begin{eqnarray*}
P^\ast(s) := \left\{\begin{array}{ll}
x(s) = y(s) = z(s)=3s| f_1|   &, \  \mbox{ $0 \le s \le \frac{1}{3}$}   \\  
x = |f_1|,  \   y(s) = z(s) = | f_1| + 3(|f_2|-|f_1|)(s-\frac{1}{3})   & , \ \mbox{$\frac{1}{3} \le s \le \frac{2}{3}$} \\
x = |f_1|,   \  y = | f_2| ,\  z(s) = | f_2| + 3(|f_3|-|f_2|)(s-\frac{2}{3})   & , \ \mbox{$\frac{2}{3} \le s \le 1$}
                \end{array} \right.
\end{eqnarray*}
which is composed of three straight lines in $xyz$ space. The complexity is given by the functional 
\begin{eqnarray*}
\mathcal{C} (\ket{\phi_T}) &=& \int_{P^\ast(s)}  \frac{ |\dot{x}| + |\dot{y}|+ |\dot{z}|  }{\sqrt{1-x(s)^2-y(s)^2 - z(s)^2}} ds  \\
&=& \sqrt{3}\  tan^{-1} \left( \frac{\sqrt{3} \ |f_1|}{\sqrt{1-3 |f_1|^2}} \right) + 
\sqrt{2} \left[ tan^{-1} \left( \frac{ \sqrt{2} |f_2|}{\sqrt{1- |f_1|^2-2|f_2|^2}} \right)
-  tan^{-1} \left( \frac{\sqrt{2} |f_1|}{\sqrt{1-3 |f_1|^2}} \right) \right]  \\
&&+ tan^{-1} \left( \frac{|f_3|}{\sqrt{1- |f_1|^2-|f_2|^2-|f_3|^2}} \right)
-  tan^{-1} \left( \frac{ |f_2|}{\sqrt{1- |f_1|^2-2|f_2|^2}} \right).
\end{eqnarray*}
The above can be generalized to any target state by finding an ordered permutation such that $|f_{p(1)}|\le|f_{p(2)}|\le|f_{p(3)}|$ and then replacing $f_i$ by $f_{p(i)}$.   
Extending the above discussions to higher $k$, we obtain a systematic way to write down the complexity for any target state $\ket{\phi_T}$ with respect to $\ket{\phi_R}$: \\
\textbf{
(1)\ Expand the target state in terms of $\ket{\tilde{q}}$, i.e. $\ket{\phi_T} = \sum_{q=0}^k f_q\ket{\tilde{q}}$. \\
(2)\ Find a permutation $p \in S_k$ such that $|f_{p(1)}|\le|f_{p(2)}| \le \cdots |f_{p(k)}|$.  \\
(3)\ Define
\begin{eqnarray*}
\mathcal{C}_1  &\equiv&  \sqrt{k}\  tan^{-1} \left( \frac{\sqrt{k} \ |f_{p(1)}|}{\sqrt{1-k |f_{p(1)}|^2}} \right)  \\
\mathcal{C}_i &\equiv&  \sqrt{k-i+1}\left[  tan^{-1} \left( \frac{\sqrt{k-i+1} \ |f_{p(i)}|}{ \sqrt{1-F_i}} \right) 
-  tan^{-1} \left( \frac{\sqrt{k-i+1} \ |f_{p(i-1)}|}{\sqrt{1-F_{i-1}}} \right) 
 \right]   \quad for \ 2 \le i \le k
\end{eqnarray*}
\quad where 
\begin{eqnarray*}
F_{i} \equiv (k-i)| f_i|^2 +  \sum_{j=1}^i |f_j|^2
\end{eqnarray*} 
\\(4)\ The complexity of the target state is defined as $\mathcal{C} (\ket{\phi_T}) = \sum_{i=1}^k \mathcal{C}_i$.
}
 
$\mathcal{C}_i$ is the complexity of the $i^{th}$ straight line in the minimal path $P^\ast(s)$. The permutation step ensures that exchanging any two coefficients $f_i,\ f_j$ for $i,j\neq 0$ does not change the complexity. 
 
Although we have constructed a notion of complexity in the context of Chern Simons theory, the above algorithm can be applied to any single qubit state (with $k$- dimensional Hilbert space). The requirement of this construction is to specify a set of physical basis states, which contain the reference state as one of its elements. Once such a basis is determined, one can naturally define $2k-2$ minimal, universal fundamental generators which can map the reference state to arbitrary target states. The complexity is then a function of the magnitudes of coefficients expanded in this basis, as described in the above algorithm.

\subsubsection{Extension to general link complement states}\label{exten}

So far we have only considered torus link complement states because their symmetry properties reduce them to an effective Hilbert space on a single site. This feature allows us to use only "small" gates and generators to prepare any torus link complement states.  Then via $CNOT$ operators, which only involve two sites, we implement the GHZ-like structure. If we eliminate the restriction of only using small generators, then such a definition can be extended to prepare any link complement state. 
 
For example, consider $n$-component link complement states in $SU(2)_k$ Chern Simons theory which can be expressed as
\begin{eqnarray*}
\ket{\mathcal{L}^n} = \sum_{q_1,\cdots,q_n} C(q_1,\cdots ,q_n)  \ket{\tilde{q}_1 ,\cdots,\tilde{q}_n}  
\end{eqnarray*}  
which is in $\mathbb{CP}^{(k+1)^n}$. Such states do not have GHZ-like structure in general. Again the reference state is chosen to be the unknot state, which takes a simple form in the basis $\{\ket{\tilde{q}}\}$:
\begin{eqnarray*}
\ket{\phi_R} =  \ket{\tilde{0} ,\cdots,\tilde{0}}  .
\end{eqnarray*}  
Similar to the single site case, one can choose $2(k+1)^n -2$ minimal universal fundamental generators to prepare any states in $\mathbb{CP}^{(k+1)^n}$ from $\ket{\phi_R}$. Such generators can be chosen as rotation generators from $\ket{\tilde{0} ,\cdots,\tilde{0}} $ to $\ket{\tilde{q}_1 ,\cdots,\tilde{q}_n}  $, for $({q}_1 ,\cdots,{q}_n) \neq (0,\cdots,0)$. Each rotation contains two generators (analogous to $\sigma_x,\sigma_y$) so the degrees of freedom match the dimension of $\mathbb{CP}^{(k+1)^n}$. Then one can compute the complexity in the same way used in the above subsections. For torus link complement states, this approach reproduces the same result obtained in previous sections, up to the term $\mathcal{C}_{CNOT}$. This difference is due to the fact that there is no need to use $CNOT$ operators to inplement the GHZ-like structure, since we allow rotation between $\ket{\tilde{0} ,\cdots,\tilde{0}} $ and $\ket{\tilde{q} ,\cdots,\tilde{q}}$. The price is to eliminate  the constraint that only "small" generators are used. 

\subsection{Complexity for some particular torus link states} \label{Picture}
 
In this section we calculate the complexity of some of the simplest torus links in $SU(2)_k$ Chern Simons and discuss their features. Recall that the modular matrices in this case are given by 
\begin{eqnarray*}
T_{mn} &=& e^{2 \pi i \frac{m(m+1)}{k+2}} \delta_{mn}  \\
S_{mn} &=& \sqrt{\frac{2}{k+2}} sin \left( \frac{\pi(m+1)(n+1)}{k+2} \right).  
\end{eqnarray*}
In particular, we consider  
\begin{itemize}
\item{Hopf link}:
This is the first non-trivial torus link with two components. The state takes the  form \cite{Witten:1988hf}:
\begin{eqnarray*}
\ket{Hoft \ link} &=& C_0\sum_{j_1,j_2} \sum_q  (STS)_{0q} \frac{S_{j_1q}S_{j_2q}}{S_{0q}} \ket{j_1 \ j_2 } \\ \nonumber
 &=& C_0 \sum_q  \frac{(STS)_{0q} }{S_{0q}} \ket{\tilde{q} \ \tilde{q} } 
\end{eqnarray*}
The above form has GHZ-like structure, as expected. Using $(ST)^3 = 1$ and $S^2=1$ one can further simplify the expression:
\begin{eqnarray}
\label{Hopf}
\ket{Hopf \ link} &=& C_0\sum_q  \frac{(T^{-1}ST^{-1})_{0q} }{S_{0q}} \ket{\tilde{q} \ \tilde{q} } \\  \nonumber
&=&C_0  \sum_q  e^{-2 \pi i(h_q + h_0)} \ket{\tilde{q} \ \tilde{q} } 
\end{eqnarray}
The normalization constant is $C_0 = \frac{1}{\sqrt{k+1}}$. We see that for the Hopf link, the magnitudes of every coefficient is the same: $|f_q| = \frac{1}{\sqrt{k+1}}$. This implies that the Hopf link state is always on the path $x=y$ suggested by the Euler-Lagrange equation above. The complexity of Hopf link can be obtained explicitly:
\begin{eqnarray}
\mathcal{C}_{Hopf} = \sqrt{k} \  cos^{-1} \left ( \frac{1}{\sqrt{k+1}} \right) 
\end{eqnarray}
For large $k$ we have $\mathcal{C}_{Hopf} \sim \sqrt{k} \cdot  \frac{\pi}{2}$. Generally in our construction, the maximal complexity can be shown to be just $ \frac{\pi}{2}\sqrt{k} $, and so for large $k$, the Hopf link state has nearly maximal complexity. In fact, it was shown in \cite{Balasubramanian:2016sro} that the Hopf link is maximally entanglement and is analogous to a Bell pair. Although there is no reason to expect that our complexity can be related to entanglement entropy, we see here a relation between maximal complexity and entanglement entropy. It will be interesting to investigate whether complexity and entropy have a deeper connection, but we leave this to future study.
\\
\item{$2N^2_1$ links}:
Next we consider a family of two component torus links, whose members are similar to the  Hopf link but the components have $2N$ crossing numbers instead, denoted as $2N^2_1$, see Fig. \ref{2N1}.
\begin{figure} [h!]
\begin{center}
\includegraphics[width=150pt]{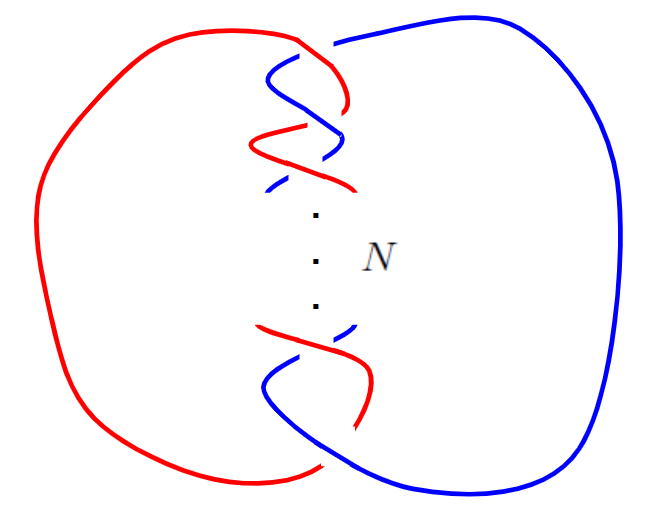} 
\caption{The $2N^2_1$ links. The Hopf link is a special case of this family with $N=1$.}
\label{2N1}
\end{center}
\end{figure}  
 $2N^2_1$ states are of the form \cite{Witten:1988hf}:   
\begin{eqnarray}
\label{TN}
\ket{2N^2_1} &=& C_0\sum_{j_1,j_2} \sum_q  (ST^NS)_{0q} \frac{S_{j_1q}S_{j_2q}}{S_{0q}} \ket{j_1 \ j_2 } \\ \nonumber
 &=& C_0 \sum_q  \frac{(ST^NS)_{0q} }{S_{0q}} \ket{\tilde{q} \ \tilde{q} } 
\end{eqnarray}
The normalized coefficients are 
\begin{eqnarray*}
f_q =  \frac{\frac{(ST^NS)_{0q} }{S_{0q}} }{ \sqrt{|\sum_j \frac{(ST^NS)_{0j} }{S_{0j}}|^2}} \end{eqnarray*}
One can solve for the complexity by the algorithm of Sec. \ref{Extension}. Using Mathematica, we have generated the complexity for $N=2$ and $N=3$ for various values of $k$ as in Figs. \ref{4_6}.
\begin{figure} [h!]
\begin{center}
\includegraphics[width=500pt]{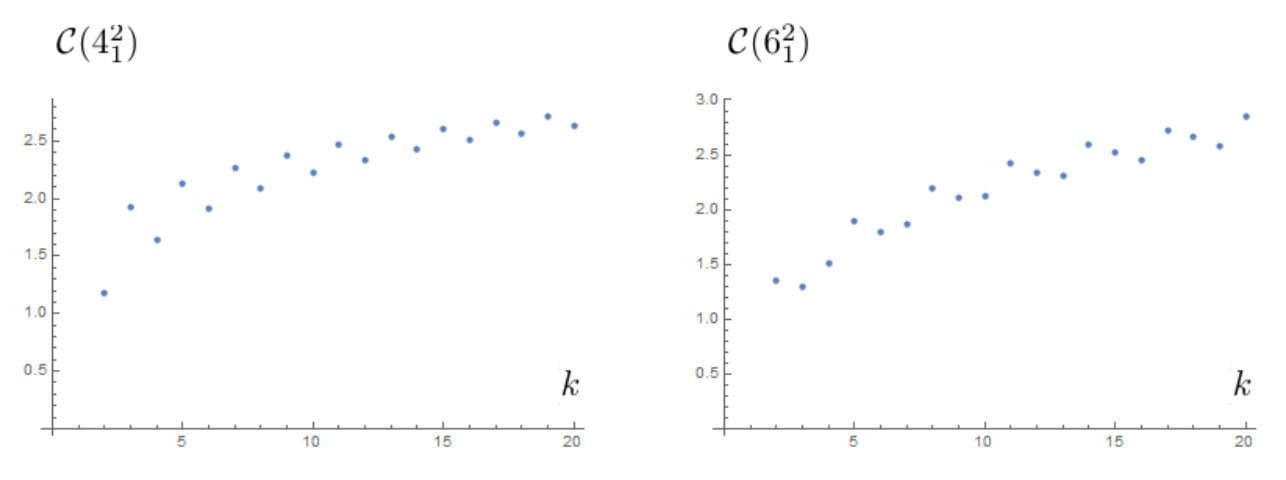} 
\caption{Left: complexity of $4^2_1$ link for various $k$. Right: complexity of $6^2_1$ link for various $k$. }
\label{4_6}
\end{center}
\end{figure}  
The complexity rises with $k$ in both cases
\footnote{The oscillation comes from the twist operator in (\ref{TN}). If we fix $N=2$ then $gcd(N,k+2)$ is different for even $k$ and odd $k$. A similar oscillation pattern also appears in the entnaglement entropy of link complement states \cite{Balasubramanian:2016sro}. }. This feature is partly due to the fact that the Hilbert space dimension is $k+1$ and the number of generators is $2k$. For larger $k$ one has "more space" to measure the complexity. In other words, the complexity defined in this paper can reflect the size of the Hilbert space. This is different from the standard inner-product distance of Hilbert space, which is always bounded by $2\pi$. 
\end{itemize}

\section{Summary and discussion} \label{Discussion} 

In this paper, we have studied the computational complexity of link complement states in Chern Simons theory. We set the reference state as the one corresponding to $n$ unlinked unknots. For the Abelian case, we find a natural set of fundamental gates, which can be thought of as raising and lowering operators of Gauss linking number between pairs of  components. The complexity is then determined by these linking numbers modulo the level $k$. This observation provides a simple and interesting connection between complexity and knot theory. In the non-Abelian case, we have chosen to focus on torus link complement states, and in that context we have shown that using their GHZ-like structure one can reduce the problem to defining the computational complexity of a single knot state. We have defined an algorithm through an extremization procedure that gives a systematic way to define complexity for these torus link complement states. 
 
In both Abelian and non-Abelian cases, we have seen that the complexity depends on the number of degrees of freedom $N$ of the system. For the $U(1)_k$ case, 
link complement states are described by $N = n(n-1)/2$ linking numbers and the maximal complexity is $\mathcal{C}_{max} \propto N$. On the other hand, for $SU(2)_k$ Chern Simons, the torus link complement states are described by $N= 2k$ independent coefficients and one has $\mathcal{C}_{max} \propto \sqrt{N}$. The two cases behave differently because they have different structure of generators. In the Abelian case, the fundamental gates are all commuting so the contribution from each can be counted separately. This fact implies that the complexity grows linearly, as we found. On the other hand in $SU(2)_k$, generators do not commute and the complexity grows more slowly. 
 
The reference state $\ket{\phi_R}$ and the basis $\{\ket{\tilde{q}}\}$ play a central role in our construction of complexity for the non-Abelian case because we choose the fundamental generators corresponding to the rotations between them. This choice breaks the homogeneity because we only use $2k$ minimal generators among the total $(k+2)(k+1)/2$ generators of the operator space $SU(k+1)$. Since $\ket{\phi_R}$ and $\{\ket{\tilde{q}}\}$ are natural physical states in the context of Chern Simons theory, our construction provides a non-trivial definition of complexity based on physical considerations. Furthermore, this definition is related to circuit complexity directly because it just counts the amount needed for each fundamental generator. If the constraint of using only "small"
generators is released, one can use the same method to define the complexity for arbitrary states.  

It would be of interest to investigate if our methods can be applied to other cases. For example, in \cite{Balasubramanian:2016sro}, link complement states with hyperbolic structure were shown to have interesting properties. It would be of interest to investigate how our notion of complexity might be related to other geometric notions of complexity in that context. We note the recent related paper \cite{Fliss:2020yrd}.

\paragraph{Acknowledgements}

We are grateful to Jackson Fliss for enlightening discussions.
This work was supported by the U.S. Department of Energy under contract 
DE-SC0019517.


\bibliographystyle{uiuchept}
\bibliography{complexity}

\end{document}